%% file: main.tex
\documentclass[runningheads]{llncs}
\usepackage[T1]{fontenc}
\usepackage{graphicx}

\usepackage[american]{babel}

\usepackage{siunitx} 
\usepackage{booktabs} 
\usepackage{tikz} 

\usepackage{xspace}
\usepackage{multirow}
\usepackage{caption}
\usepackage{subcaption}
\RequirePackage{algorithm}
\RequirePackage{algorithmic}
\usepackage{enumitem}
\usepackage{wrapfig}
\usepackage{hyperref}

\usepackage{amssymb}
\usepackage{pifont}
\newcommand{\cmark}{\ding{51}}%
\newcommand{\xmark}{\ding{55}}%

\input{definitions.tex}

\usepackage{lipsum}

\newcommand{\abstain}{{\ensuremath{\oslash}}\xspace}

\usepackage{color}

\begin{document}

\title{Certification of Deep Learning Models for Medical Image Segmentation}

\titlerunning{Certification of Medical Image Segmentation Models}

\author{Othmane Laousy\inst{1,2,3} \and
Alexandre Araujo\inst{4} \and
Guillaume Chassagnon\inst{2} \and
Nikos Paragios\inst{5} \and
Marie-Pierre Revel\inst{2} \and
Maria Vakalopoulou\inst{1,3}}


%
\authorrunning{O. Laousy et al.}
%
\institute{MICS, CentraleSupélec, Paris-Saclay University, Gif-sur-Yvette, France.
\and
Hôpital Cochin, AP-HP, Paris-Cité University, Paris, France
\and
Inria Saclay, Gif-sur-Yvette, France
\and
New York University, NY, USA
\and
Therapanacea, Paris, France
}

\maketitle              

\begin{abstract}
In medical imaging, segmentation models have known a significant improvement in the past decade and are now used daily in clinical practice. However, similar to classification models, segmentation models are affected by adversarial attacks. In a safety-critical field like healthcare, certifying model predictions is of the utmost importance. Randomized smoothing has been introduced lately and provides a framework to certify models and obtain theoretical guarantees. 
In this paper, we present for the first time a certified segmentation baseline for medical imaging based on randomized smoothing and diffusion models. 
Our results show that leveraging the power of denoising diffusion probabilistic models helps us overcome the limits of randomized smoothing. We conduct extensive experiments on five public datasets of chest X-rays, skin lesions, and colonoscopies, and empirically show that we are able to maintain high certified Dice scores even for highly perturbed images. 
Our work represents the first attempt to certify medical image segmentation models, and we aspire for it to set a foundation for future benchmarks in this crucial and largely uncharted area.

\end{abstract}

\section{Introduction}

For the past decade, deep neural networks have dominated the computer vision community and provided near human performance on many different tasks, including classification~\cite{he2016deep}, segmentation~\cite{long2015fully}, and image generation~\cite{goodfellow2020generative}.
Given these impressive results, convolutional neural networks are now used on a daily basis in fields like healthcare, self-driving cars, and robotics, to cite a few.
In medical imaging, convolutional neural networks are particularly used to segment organs or regions of interest on different modalities such as X-rays, CT scans, MRIs, or ultrasound~\cite{lei2020medical}.
Indeed, segmentation techniques and variations of 2D and 3D U-Nets are currently the state-of-the-art to identify and isolate tumors, blood vessels, organs, or other structures within an image and provide crucial help to physicians for medical diagnosis, screening, and prognosis~\cite{unetreview}.

Nowadays, segmentation models are gaining widespread adoption in modern clinical practice and are being used with increasing frequency, making the results of these models critical for many patients.
However, it is now commonly known that neural networks can be vulnerable to adversarial attacks~\cite{goodfellow2014explaining,szegedy2013intriguing}, \ie, small input perturbations invisible to humans crafted specifically such that the network performs errors.
Over the past few years, a large body of work has devised empirical defenses against adversarial attacks for classification tasks~\cite{goodfellow2014explaining,madry2017towards,araujo2020advocating}, as well as segmentation tasks~\cite{xie2017adversarial}, including applications on medical imaging~\cite{paschali2018generalizability}.
Although state-of-the-art empirical defenses provide significant robustness, these defenses do not guarantee {\em theoretical} robustness and stronger attacks can be crafted to break them~\cite{athalye2018obfuscated}.
Recently, {\em certified} defenses, for classification~\cite{cohen2019certified,meunier2022dynamical,araujo2023a} and segmentation~\cite{fischer2021scalable,laousy2023towards}, have been proposed to guarantee the accuracy and reliability of neural networks.
However, certified defenses for segmentation in the context of medical imaging are still lacking, even if models are getting market approvals (\eg, FDA, CE) and are already adopted in clinical practice.

In this paper, we provide the first method for certified robustness in the context of segmentation for medical imaging.
We leverage the {\em randomized smoothing} strategy~\cite{cohen2019certified,fischer2021scalable}, and the recent work on {\em diffusion models}~\cite{carlini2023certified} to achieve state-of-the-art certified robustness for segmentation models.
Randomized smoothing consists in convolving the neural network with a Gaussian distribution (\ie, by adding noise to the input) in order to obtain a smooth segmentation model.
From the smoothness properties of the segmentation model, we can derive a robustness guarantee and compute a certified Dice score.
We go even further by using diffusion models to first denoise the perturbed input and boost the certified robustness.
By extension, we show that current diffusion models, trained on `classical images' generalize well to medical datasets for denoising tasks. Extensive experiments
on five public medical datasets of chest X-rays~\cite{shiraishi2000development,jaeger2014two}, skin lesions~\cite{skin}, and colonoscopies~\cite{bernal2015wm}, and different popular segmentation models, prove the potential of our method. We hope that this study will provide the first step towards robustness guarantees for medical image segmentation.

\section{Related Work}
Since the discovery of adversarial attacks~\cite{szegedy2013intriguing,goodfellow2014explaining}, numerous defenses~\cite{madry2017towards,goodfellow2014explaining,carlini2017towards} and attacks have been devised~\cite{carlini2017towards,madry2017towards}, demonstrating that neural networks are sensitive to small input perturbation and vulnerable to attacks.
Adversarial training, which has been acknowledged as one of the most successful empirical defenses, consists in training a network directly on adversarial examples~\cite{madry2017towards}.
However, it is now known that even strong defenses can be bypassed by adaptive attacks~\cite{croce2020reliable}.
Paschali et al. \cite{paschali2018generalizability} were among the first to study adversarial attacks in the context of medical imaging.
They conducted experiments using several neural network architectures~\cite{szegedy2017inception,howard2017mobilenets} (\ie, Inception V3, V4, MobileNet) and several attacks~\cite{goodfellow2014explaining,madry2017towards} to demonstrate that the vulnerability of neural networks is extended to medical images.

More specifically, in the context of classification, a previous work \cite{asgari2018vulnerability} has analyzed the robustness of neural networks for chest X-ray images and showed that gradient-based attacks were successful in fooling both machines and humans.
In a similar line of work, Yao et al. \cite{yao2021hierarchical} proposed an add-on to known attacks that bypasses state-of-the-art adversarial detectors making current defenses even less robust.
On the other hand, several works have been focused on crafting defense strategies specifically in the context of medical imaging.
For example, Almalik et al.~\cite{almalik2022self} proposed a self-ensembling method to enhance the robustness of Vision Transformers in the presence of adversarial attacks.
In the context of segmentation in medical imaging,~\cite{santhirasekaram2022vector} introduced a vector quantization approach by learning a discrete representation in a low dimensional embedding space and improving the robustness of a segmentation model.
Finally, Daza et al.~\cite{daza2021towards} proposed a lattice architecture that segments organs and lesions on MRI and CT scans and leveraged an efficient approach of adversarial training to defend against adversarial examples. 

Although a large body of work has focused on constructing defenses for classification and segmentation tasks in the context of medical imaging, {\em certified} defenses are under-studied by the medical community.
In this paper, we propose to leverage randomized smoothing and diffusion models for certified segmentation on medical datasets, setting the first baseline for this challenging problem and certifying popular segmentation architectures.

\section{Randomized Smoothing}

Randomized smoothing is a model agnostic technique, proposed by Cohen et al.~\cite{cohen2019certified}, used to improve and certify the robustness of neural networks against adversarial attacks.
This method consists in adding random noise (\eg, noise generated from a Gaussian distribution) to the input data and then classifying the perturbed data using the neural network.
Let $\Dcal = \Xcal \times \Ycal$ denote the data distribution where $\Xcal \subset \Rbb^d$ and $\Ycal = \{ 1, \dots, k \}$ represent the input space and target space respectively and $k$ is the number of classes.
Let $f: \Xcal \rightarrow \Ycal$ be a neural network such that for $(x, y) \in \Dcal$, the classifier correctly classifies if $f(x) = y$.
An adversarial attack is a small norm-bounded perturbation $\delta \in \Rbb^d$ with $\norm{\delta}_2 \leq \epsilon$ such that: $f(x + \delta) \neq y$.
Randomized smoothing is a procedure to construct a new {\em smooth} classifier $g$ given any base classifier $f$. Let $\Ncal(0, \sigma^2 \Imat)$ be a Gaussian distribution of mean $0$ and variance $\sigma$, then, the smooth classifier $g$ is defined as follows:
\begin{equation*}
  g(x) = \Pbb_{\eta \sim \Ncal(0, \sigma^2 \Imat) } \left[ f(x + \eta) = y \right]
\end{equation*}
Cohen et al.~\cite{cohen2019certified} have shown that if $R = \sigma\Phi^{-1}(g(x))$ where $\Phi$ is the cumulative distribution function of the standard Gaussian distribution and $R$ can be considered the certified radius, then, $g(x + \delta) = y$ for all $\delta$ satisfying $\norm{\delta}_2 \leq R$.
However, since it is not possible to compute $g$ at $x$ exactly, they proposed using Monte Carlo algorithms as an alternative approach for estimating $g(x)$ using random sampling.
In order to obtain a reliable estimate of the probability $g(x)$, they also suggested a method that involves generating $n$ samples of $\eta$ from a normal distribution $\Ncal(0, \sigma^2 \Imat)$ and evaluating $f(x + \eta)$ for each sample.
The resulting counts for each class in $\Ycal$ are then used to estimate probability $p_y$ and the radius $R$ with confidence $1-\alpha$ (where $\alpha$ is a value between 0 and 1).
If the confidence level cannot be achieved (for example, due to insufficient samples), the method will abstain from providing an estimate.
More recently, Fischer et al.~\cite{fischer2021scalable} built upon the work of~\cite{cohen2019certified} by introducing \textsc{SegCertify}, the first certified approach for image segmentation.
The segmentation process involves assigning a segmentation class to every pixel in the image, which can be viewed as a form of classification at the pixel level.
In the segmentation settings, the output space consists of regions or categories to be segmented, such as cars, roads, pedestrians, etc.
The classifier function $f: \Xcal \rightarrow \Ycal^{d}$ determines the class for each pixel and categorizes each component independently.
In this context, the certification algorithm proposed by Cohen et al.~\cite{cohen2019certified} can be extended to accommodate the segmentation task.

To obtain a smooth classifier, it is necessary to add random noise to the input of the classifier. However, this creates a trade-off between accuracy and robustness.
If low variance noise is added, accuracy won't be impacted significantly, but the certified radius will remain low.
Conversely, adding high variance noise can improve certificates but at the expense of accuracy.
To address this issue, Cohen et al. proposed a simple trick of training the network with noise injection during the training phase.
While this method may reduce accuracy when evaluating the classifier with noise during the certification process, it can also help mitigate the trade-off between accuracy and robustness.
One can note that during training, the network's objective is to learn to ignore the noise and classify at the same time. 
To improve the natural as well as the certified accuracy, Salman et al.~\cite{salman2020denoised} proposed to separate the two tasks with two networks trained separately.
First, a network, $h: \Xcal \rightarrow \Xcal$, is trained to denoise the data such that for $\eta \sim \Ncal(0, \sigma^2 \Imat)$, we have $h(x + \eta) \approx x$, then, the output of the denoiser is given to the classifier.

In this paper, we leverage randomized smoothing and diffusion probabilistic models to obtain state-of-the-art results on certified segmentation for medical imaging.
To the best of our knowledge, we are the first to propose a comprehensive study on certified segmentation for medical imaging.

\section{Diffusion Probabilistic Models for Certification}

The training of a Denoising Diffusion Probabilistic Model (DDPM) is an iterative process that involves adding a small amount of noise at every step of the diffusion process until random noise is reached.
The reverse process then starts from random noise and generates a new image that conforms to the data distribution.
Since DDPMs are inherently iterative denoising models, we can leverage this property for randomized smoothing.
The idea would be to start the reverse process with a noisy image, rather than Gaussian noise, enabling the DDPM to output an image that resembles the original image. 

Similar to Carlini et al.\cite{carlini2023certified}, our proposed pipeline is composed of two main steps: we denoise, then we certify. 
In order to complete the denoising, we need to first map between the noise model utilized in diffusion models and the one used in randomized smoothing. Randomized smoothing needs a data point that is enhanced with Gaussian noise added to it, given by $x_\text{rs} = x + \delta$ with $\delta \sim \Ncal(x, \sigma^2 \Imat)$. On the other hand, diffusion models suppose the noise model for $x_\text{DDPM} \sim \Ncal(\sqrt{\alpha_t}x, (1-\alpha_t)\Imat)$.
Programmatically, we start by adding Gaussian noise to an image $x$, obtaining $x_\text{rs}$. Then the timestep $t^*$ on which we can use the diffusion model for randomized smoothing is defined. Depending on the scheduler of the denoiser, we compute $t^*$ such that $\sigma^2 = \frac{1-\alpha_{t^*}}{\alpha_{t^*}}$ (obtained by scaling $x_\text{rs}$ with $\sqrt{\alpha_t}$ and pairing the variances). We then calculate $x_\text{DDPM} = \sqrt{\alpha_{t^*}} ( x + \delta), \delta \sim \Ncal(0, \sigma^2 \Imat)$. After that, we apply a single-step denoiser and predict the completely denoised image. A single-step denoising involves directly predicting the image from $t^*$ to $t=0$. A multi-step denoising strategy implies iteratively predicting all images at $t^*$, $t^*-1$, $\dots$ until $t=0$. Both techniques are explored in the next section and supplementary material.

Since randomized smoothing is applied to each pixel separately with a probability of $1 - \alpha$, considering the entire segmentation region would imply considering a union bound with significantly reduced confidence. Similar to Fischer et al.~\cite{fischer2021scalable}, we leverage the Holm-Bonferroni method~\cite{holm1979simple} and perform multiple-testing corrections. For each image, we repeat this process $n=100$ times, identifying pixels on which the model abstains, and computing the certified scores.
We extend the work of Fischer et al.~\cite{fischer2021scalable} to also compute a certified Dice score that is calculated ignoring the abstain class ($\abstain$).
Our approach has a significant advantage compared to \textsc{SegCertify} since it leverages off-the-shelf and state-of-the-art pre-trained denoisers and segmentation models. \textsc{SegCertify}, on the other hand, relies on models trained with Gaussian noise.

\begin{table} 
\centering
{\small
\caption{Comparison of our approach with three different model architectures on chest X-ray datasets. We report certified Dice, IoU and percentage of abstentions (\%\abstain) for different noise levels \boldsymbol{$\sigma$} and radii \boldsymbol{$R$}.}\label{table:denoise}
\begin{tabular}{lcccccccccccccccccc}\toprule
& & & \multicolumn{7}{c}{\textbf{JSRT}} & &\multicolumn{3}{c}{\textbf{Montgomery}} & &\multicolumn{3}{c}{\textbf{Schenzen}} \\
\cmidrule{4-10}\cmidrule{12-14}\cmidrule{16-18}
\hfil\multirow{2}{*}{\boldsymbol{$\sigma$}}\hfill &\multirow{2}{*}{\boldsymbol{$R$}} & &\multicolumn{2}{c}{Lung}                         &\multicolumn{2}{c}{Heart} &\multicolumn{2}{c}{Clavicles} & & &\multicolumn{3}{c}{Lung} & &\multicolumn{3}{c}{Lung} \\                   
\cmidrule{4-10}\cmidrule{12-14}\cmidrule{16-18}
 & & & Dice &IoU &Dice &IoU &Dice &IoU &\%\abstain & &Dice &IoU &\%\abstain & &Dice &IoU &\%\abstain \\
\midrule
\multicolumn{5}{l}{\textbf{UNet~\cite{ronneberger2015u}}} & & & & & & & & & & & &  \\  
\midrule
0.25\phantom{aa} &0.17 & \phantom{aa} &0.94 &0.91 &0.88 &0.79 &0.75 &0.63 &0.07 & &0.93 &0.89 &0.07 & &0.95 &0.90 &0.05 \\
0.50 &0.34 &&0.90 &0.83 &0.88 &0.79 &0.61 &0.45 &0.09 & &0.89 &0.80 &0.07 & &0.93 &0.90 &0.02 \\
1.00 &0.67 &&0.87 &0.79 &0.84 &0.75 &0.23 &0.15 &0.15 & &0.88 &0.80 &0.14 & &0.89 &0.83 &0.10 \\
\midrule
\multicolumn{5}{l}{\textbf{ResUNet++~\cite{jha2019resunet}}} & & & & & & & & & & & & \\
\midrule
0.25 &0.17 &&0.95 &0.91 &0.93 &0.87 &0.78 &0.65 &0.05 & &0.96 &0.93 &0.02 & &0.95 &0.91 &0.01 \\
0.50 &0.34 &&0.94 &0.88 &0.91 &0.83 &0.63 &0.48 &0.08 & &0.94 &0.89 &0.03 & &0.93 &0.90 &0.02 \\
1.00 &0.67 &&0.90 &0.82 &0.87 &0.77 &0.28 &0.19 &0.12 & &0.89 &0.83 &0.07 & &0.90 &0.85 &0.06 \\
\midrule
\multicolumn{5}{l}{\textbf{DeeplabV2~\cite{deeplab}}} & & & & & & & & & & & & \\
\midrule
0.25 &0.17 &&0.94 &0.91 &0.91 &0.86 &0.85 &0.75 &0.04 & &0.93 &0.91 &0.07 & &0.80 &0.71 &0.07 \\
0.50 &0.34 &&0.88 &0.81 &0.87 &0.79 &0.63 &0.49 &0.10 & &0.91 &0.87 &0.02 & &0.34 &0.25 &0.15 \\
1.00 &0.67 &&0.88 &0.80 &0.83 &0.74 &0.20 &0.11 &0.14 & &0.85 &0.79 &0.17 & &0.04 &0.02 &0.11 \\
\bottomrule
\end{tabular}
}
\end{table}

\section{Experiments and Results}

\textbf{Datasets:}
We perform experiments on $5$ different publicly available datasets. All datasets were divided to $70\%$ for training, $10\%$ for validation, and $20\%$ for testing. The testing set is the one used to compute certified results. 

\textbf{Chest X-rays datasets:} JSRT dataset~\cite{shiraishi2000development} with annotations of lung, heart, and clavicles provided by~\cite{jsrtannot} is used. This dataset contains $247$ images.
For lung segmentation only, we use both the Montgomery and Shenzen datasets \cite{jaeger2014two}. Montgomery consists of 138  and Shenzen of 662 annotated images. 

\textbf{Skin lesion:} Skin images with their annotations provided by the ISIC 2018 boundary segmentation challenge~\cite{skin} were used. This dataset consists of $2694$ RGB dermatoscopy images. 

\textbf{Colonoscopy images:} CVC-ClinicDB dataset \cite{bernal2015wm} containing $612$ colonoscopy images in RGB together with their annotations were utilized.

\textbf{Implementation Details:}
We train three different segmentation models namely, a UNet\cite{ronneberger2015u}, a ResUNet++\cite{jha2019resunet}, and a DeeplabV2\cite{deeplab} with and without noise. The models trained without noise are used exclusively with our method. The models trained with a Gaussian noise of $0.25$ are used to compute \textsc{SegCertify} scores. All 6 models use an image input size of $512\times512$ for X-ray images, $384\times512$ for skin lesions, and $288\times384$ for colonoscopy.
As a denoiser, we use an off-the-shelf denoising diffusion probabilistic model provided by \cite{dhariwal2021diffusion}. We perform our experiments with the $256\times256$ class unconditional denoiser with a linear scheduler and without timestep respacing. For each noise level, our method follows the steps described in the previous section and uses $n_0=10$, n=$100$ for each image, and $\alpha=0.001$, and $\tau=0.75$. Our code is made publicly available at: \url{https://github.com/othmanela/medical_cert_seg}

\textbf{Results and Discussion:}
For all five datasets, we compute a certified Dice score and certified mean Intersection over Union (IoU). We also report the percentage of abstentions (\%\abstain) representing the mean number of pixels on which the model's prediction confidence was insufficient with respect to the radius $R$. The lower the percentage of abstentions the better the segmentation model is.

\begin{table} 
\centering
\caption{Certified segmentation results of our technique and \textsc{SegCertify}~\cite{fischer2021scalable} on the chest X-ray JSRT dataset. We report Dice, IoU, and percentage of abstentions (\%\abstain) for each class.}\label{table:jsrt}
{\small
\begin{tabular}{lcccccccccccccccc}\toprule
\multirow{2}{*}{\textbf{Model}} &\multirow{2}{*}{\shortstack{\textbf{Trained} \\ \textbf{with noise}}}  & &\multirow{2}{*}{\boldsymbol{$\sigma$}} & &\multirow{2}{*}{\boldsymbol{$R$}} & &\multicolumn{2}{c}{\textbf{Lung}} & &\multicolumn{2}{c}{\textbf{Heart}} & &\multicolumn{2}{c}{\textbf{Clavicles}} & \\\cmidrule{8-9}\cmidrule{11-12}\cmidrule{14-15}
& & & & & & &Dice &IoU & &Dice &IoU & &Dice &IoU &\%\abstain \\ 
\midrule
\multirow{2}{*}{ResUNet++\cite{jha2019resunet}} & \xmark & &0.00 & &0.00 & &0.97 &0.94 & &0.94 &0.91 & &0.93 &0.91 &0.00 \\
& \cmark & &0.25 & &0.00 & &0.91 &0.90 & &0.89 &0.87 & &0.84 &0.79 &0.00 \\
\midrule
\multirow{3}{*}{\textsc{SegCertify\cite{fischer2021scalable}}} & \cmark & &0.25 & &0.17 & & \textbf{0.96} & \textbf{0.92} & & \textbf{0.93} & \textbf{0.88} & & \textbf{0.83} & \textbf{0.72} & \textbf{0.04} \\
& \cmark & &0.50 & &0.34 & &0.89 &0.84 & &0.85 &0.79 & &0.58 &0.43 &0.13 \\
& \cmark & &1.00 & &0.67 & &0.07 &0.04 & &0.02 &0.01 & &0.00 &0.00 &0.24 \\
\midrule
\multirow{3}{*}{\shortstack[l]{Ours}} & \xmark & &0.25 & &0.17 & &0.95 &0.91 & & \textbf{0.93} &0.87 & &0.78 &0.65 &0.05 \\
& \xmark & &0.50 & & 0.34 & & \textbf{0.94} & \textbf{0.88} & & \textbf{0.91} & \textbf{0.83} & & \textbf{0.63} & \textbf{0.48} & \textbf{0.08} \\
& \xmark & &1.00 & &0.67 & & \textbf{0.90} & \textbf{0.82} & & \textbf{0.87} & \textbf{0.77} & & \textbf{0.28} &\textbf{0.19} & \textbf{0.12} \\
\bottomrule
\end{tabular}
}
\end{table}

In Table \ref{table:denoise}, we compare our method using $3$ different and popular architectures (UNet, ResUNet++, and DeeplabV2) on the chest X-rays datasets. We notice that our method maintains overall good results on all three model backbones. A similar table with \textsc{SegCertify} results is provided in Table S$2$ of the supplementary material. Overall, for both methods, ResUNet++ is the most robust architecture followed by UNet and then DeeplabV2 for all \boldsymbol{$\sigma$} and \boldsymbol{$R$} combinations.  Moreover, certified metrics for lungs and heart remain high for our method, even with high levels of noise. However, the increasing level of noise affects the clavicles since these are smaller structures.

A comparison of our method and \textsc{SegCertify} using the ResUNet++ architecture is presented  in Table \ref{table:jsrt} for the three chest X-ray datasets. We observe that we outperform \textsc{SegCertify}, especially for high sigma values. For $\sigma=0.25$, \textsc{SegCertify} performs slightly better. This is due to the fact that the model used with \textsc{SegCertify} is trained with a noise level of 0.25. The main drawback however is that its Dice on unperturbed images drops considerably (\eg, from 0.96 to 0.91 on lung segmentation). On the other hand, our pipeline does not require training a segmentation model with noise or even a denoising model. Our methodology relies only on off-the-shelf models. For the highest noise level of $\sigma=1.0$, we notice that the certified Dice and IoU with \textsc{SegCertify} both drop to $0$ whereas our proposed method is able to maintain high certified scores.

\begin{figure*}[t!]
  \centering
  \hfill
  \begin{subfigure}[h]{\textwidth}
    \centering
    \includegraphics[width=0.22\textwidth]{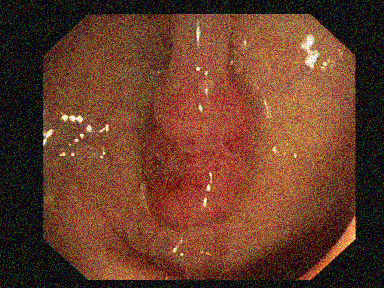}
    \includegraphics[width=0.22\textwidth]{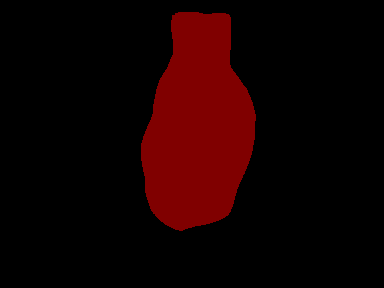}
    \includegraphics[width=0.22\textwidth]{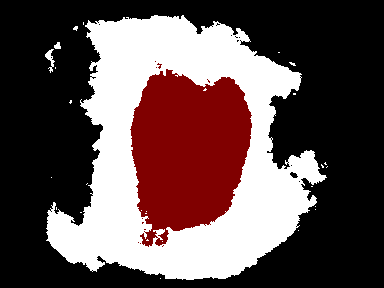}
    \includegraphics[width=0.22\textwidth]{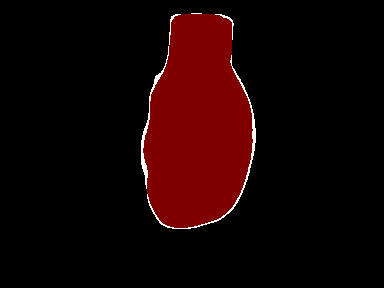}
  \end{subfigure} \\
  \begin{subfigure}[h]{\textwidth}
    \centering
    \includegraphics[width=0.22\textwidth]{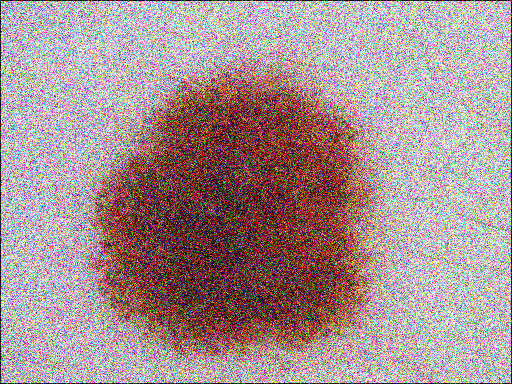} 
    \includegraphics[width=0.22\textwidth]{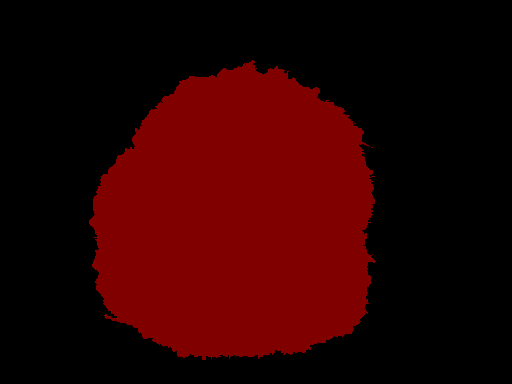}
    \includegraphics[width=0.22\textwidth]{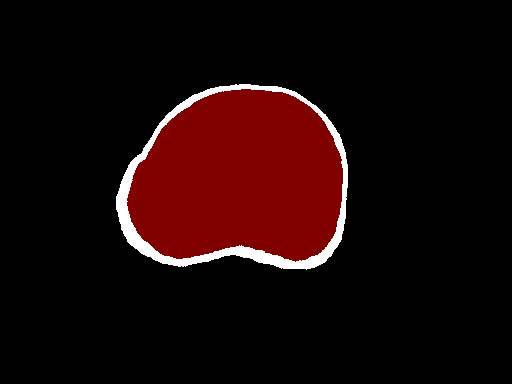} 
    \includegraphics[width=0.22\textwidth]{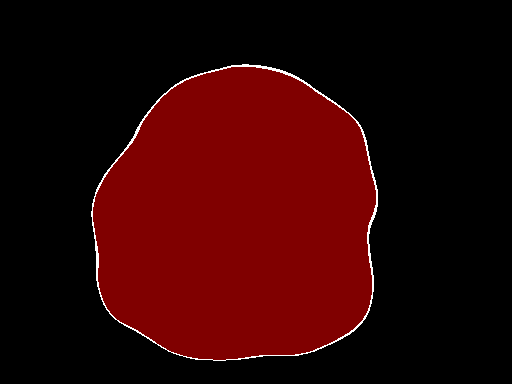}
  \end{subfigure} \\
  \begin{subfigure}[h]{\textwidth}
    \centering
    \includegraphics[width=0.22\textwidth]{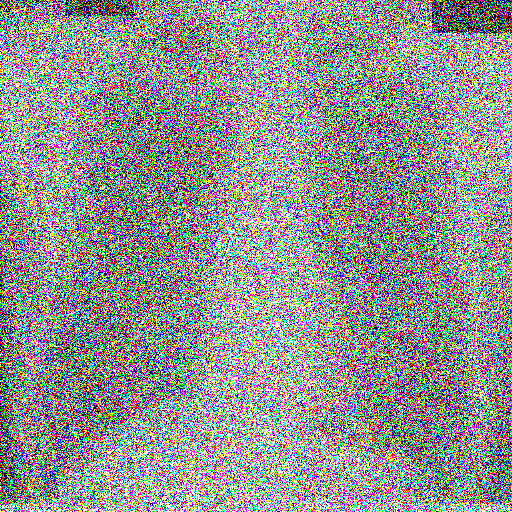}
    \includegraphics[width=0.22\textwidth]{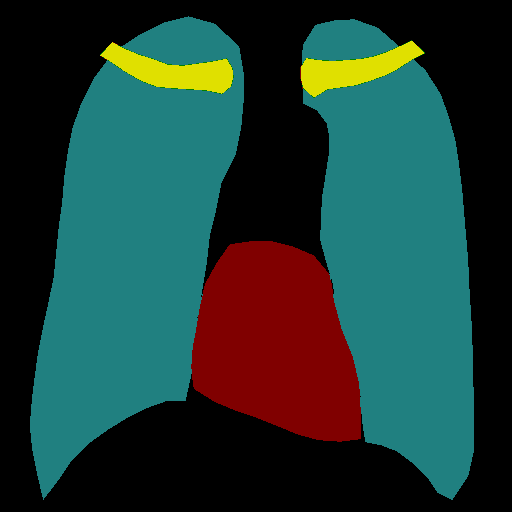}
    \includegraphics[width=0.22\textwidth]{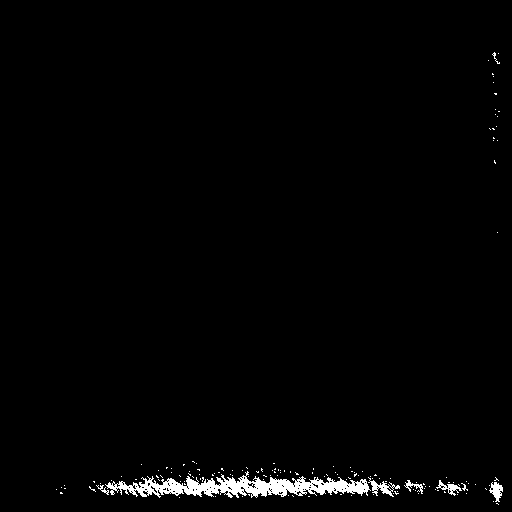}
    \includegraphics[width=0.22\textwidth]{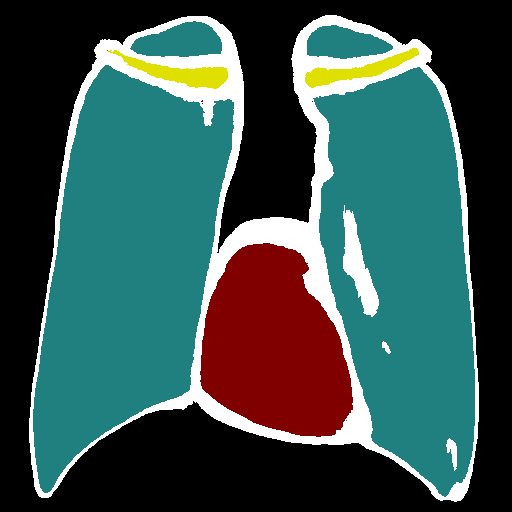}
  \end{subfigure} \\
  \begin{subfigure}[h]{0.24\textwidth}
    \centering
  \end{subfigure}
  \caption{Qualitative results of \textsc{SegCertify} and our method on colonoscopy, skin lesion, and chest X-ray images. From left to right: image with added noise, ground truth, \textsc{SegCertify} segmentation, our segmentation. White pixels denote abstention areas of the segmentation models. We increase the noise level from top to bottom: $\sigma = 0.25, 0.5$, and $1.0$. }
  \label{figure:examples}
\end{figure*}

Qualitative results are provided in Figure \ref{figure:examples} for our proposed method and \textsc{SegCertify} for the different datasets and different levels of noise. Regarding the structures to segment, we notice that the abstentions around the clavicles (the smallest benchmarked region of interest on chest X-rays) get bigger. We also notice that the fine segmentation boundaries (\eg, area around the skin lesion) may not be as sharp after denoising. As we increase the noise, the decision boundary is harder to find for all models. This may be due to the fact that fine details on the image are lost after the denoising step. However, our method is still able to segment the large majority of pixels properly on the image, contrary to its competitor, especially for high noise levels (third row on chest X-rays).

\begin{table} 
\centering
{\small
\caption{Results on skin lesions~\cite{skin} and CVC-ClinicDB~\cite{bernal2015wm} segmentation.}
\label{table:skinpolyp}
\begin{tabular}{llccccccccccc}\toprule
\multirow{2}{*}{\textbf{Model}} & \multirow{2}{*}{\textbf{Method}} &\multirow{2}{*}{\boldsymbol{$\sigma$}} &\multirow{2}{*}{\boldsymbol{$R$}} & &\multicolumn{3}{c}{\textbf{Skin Lesions}} & &\multicolumn{3}{c}{\textbf{CVC-ClinicDB}} \\
\cmidrule{6-8}\cmidrule{10-12}
& & & & &Dice &IoU &\%\abstain & &Dice &IoU &\%\abstain \\
\midrule
\multirow{6}{*}{ResUNet++\cite{jha2019resunet}\phantom{aa}} 
&\multirow{3}{*}{\shortstack[l]{\textsc{SegCertify\cite{fischer2021scalable}}\phantom{aa}}} &0.25 &0.17 & &0.79 &0.68 &0.07 & &0.63 &0.56 &0.05 \\
&&0.50 &0.34 & &0.41 &0.27 &0.06 & &0.15 &0.10 &0.01 \\
&&1.00 &0.67 & &0.00 &0.00 &0.01 & &0.00 &0.00 &0.00 \\
\cmidrule{2-12}
&\multirow{3}{*}{\shortstack[l]{Ours}} 
&0.25 &0.17 & &0.85 &0.77 &0.03 & &0.65 &0.57 &0.04 \\
&&0.50 &0.34 & &0.83 &0.76 &0.04 & &0.45 &0.39 &0.07 \\
&&1.00 &0.67 & &0.77 &0.69 &0.06 & &0.26 &0.23 &0.14 \\
\bottomrule
\end{tabular}
}
\end{table}

Table \ref{table:skinpolyp} reports certified segmentation results for skin lesions and colonoscopy on both techniques. We notice that our method is still performing better than \textsc{SegCertify}. This supports our claim that DDPMs generalize quite well to medical images and that harnessing their potential boosts the state-of-the-art. Regarding the denoiser, we used a single-step denoising strategy, \ie, we perform a single call to the DDPM to compute the denoised image from $t^*$ to $t=0$. Another strategy could be to iteratively denoise from $t^*$, $t^{*}-1$, ... until $t=0$. However, this implies predicting a denoised image multiple times and in the end, may result in images with unwanted artifacts. We perform multi-step denoising experiments and report results in Table S$1$ of the supplementary material. We note that the single-step denoising performs best since it relies more on the denoising power of DDPMs rather than their generative capabilities, and is also faster than the multi-step approach. Finally, we perform a comparison with another denoiser architecture. We train three UNet models (one for each noise level) on the JSRT dataset. We report results in Table S3 and notice that even with custom-trained denoisers, the DDPM outperforms the UNet denoising architecture.
A comparison of denoised images is provided in Figure S1. We notice that the DDPM is able to keep high-fidelity images compared to the UNet and is therefore more relevant for certified medical image segmentation.

\section{Conclusion}

In this paper, we present the first work on certified segmentation for medical imaging, and extensively evaluate it on five different datasets and three deep learning segmentation models. 
Our technique leverages off-the-shelf denoising and segmentation models and provides the highest certified Dice and mIoU on multi-class and binary segmentation of five different datasets. With that, we are able to remove the overhead of having to train and fine-tune models specifically for robustness. This paradigm shift alleviates the dilemma of having to choose between highly accurate segmentation models or models robust to attacks.
We hope that this work serves as a baseline for the unexplored yet critical topic of certified segmentation in medical imaging.
Future work will involve extending our approach to 3D medical imaging modalities as well as exploring the realm of certified classification.

\subsubsection*{Acknowledgements.} This work was granted access to the HPC resources of IDRIS under the allocation 2023-AD011013308R1 made by GENCI and it was partially supported by the ANR Hagnodice ANR-21-CE45-0007.

\bibliographystyle{splncs04}
\bibliography{bibliography}

\end{document}

%% file: definitions.tex
\usepackage{xspace}

\usepackage{mathtools}      
\usepackage{amssymb}
\usepackage{mathdots}       
\usepackage{amsfonts}       
\usepackage{nicefrac}       
\usepackage{physics}        
\usepackage{bm}             
\usepackage{centernot}      
\usepackage{thmtools}       

\DeclareMathOperator*{\Pbb}          {\mathbb{P}}

\makeatletter
\DeclareRobustCommand\onedot{\futurelet\@let@token\@onedot}
\def\@onedot{\ifx\@let@token.\else.\null\fi\xspace}

\def\eg{\emph{e.g}\onedot} 
\def\ie{\emph{i.e}\onedot}

\makeatother

\newcommand{\leftmatrix}       {\begin{pmatrix}}
\newcommand{\rightmatrix}      {\end{pmatrix}}
\newcommand{\leftmatrixsmall}  {\begin{psmallmatrix}}
\newcommand{\rightmatrixsmall} {\end{psmallmatrix}}

\def\ddefloop#1{\ifx\ddefloop#1\else\ddef{#1}\expandafter\ddefloop\fi}

\def\ddef#1{\expandafter\def\csname #1bb\endcsname{\ensuremath{\mathbb{#1}}}}
\ddefloop ABCDEFGHIJKLMNOPQRSTUVWXYZ\ddefloop

\def\ddef#1{\expandafter\def\csname #1cal\endcsname{\ensuremath{\mathcal{#1}}}}
\ddefloop ABCDEFGHIJKLMNOPQRSTUVWXYZ\ddefloop

\def\ddef#1{\expandafter\def\csname #1mat\endcsname{\ensuremath{\mathbf{#1}}}}
\ddefloop ABCDEFGHIJKLMNOPQRSTUVWXYZ\ddefloop

\def\ddef#1{\expandafter\def\csname #1matsf\endcsname{\ensuremath{\mathsf{#1}}}}
\ddefloop ABCDEFGHIJKLMNOPQRSTUVWXYZ\ddefloop

\def\ddef#1{\expandafter\def\csname #1vec\endcsname{\ensuremath{\mathbf{#1}}}}
\ddefloop abcdefghijklmnopqrstuvwxyz\ddefloop

\usepackage{todonotes}

\newcommand{\comment}[1]{}